\documentclass[amsmath,trackchanges,twocolumn]{aastex631}
\usepackage{color,natbib,epsf,amsmath,amssymb,graphicx,bm}

\begin{document}

\title{Extra Invariant in Magnetohydrodynamics of Planets and Stars}
\author{Alexander M. Balk}
\affiliation{Department of Mathematics, University of Utah, Salt Lake City, UT 84112}
\shorttitle{Extra Invariant in Magnetohydrodynamics of Planets and Stars}
\shortauthors{Balk}
\email{balk@math.utah.edu}
\begin{abstract}\noindent
The paper establishes extra invariant in magnetohydrodynamics (MHD) of rotating and stratified fluid layer. This invariant is conserved adiabatically, i.e. approximately over long time. The existence of the invariant is interesting by itself, as such invariants are extremely rare. However, in addition, this invariant appears  to be connected to the famous dynamo phenomenon.
\end{abstract}

\section{Introduction}
\label{Sect:Intro}

One of the biggest problems in MHD is to explain the dynamo phenomenon. It turns out that MHD possesses an adiabatic invariant, that appears to imply the dynamo.
 
First, let us discuss what kind of extra conservation precisely, we are talking about. If a physical system is weakly nonlinear, the main interactions are often three-wave resonances
\begin{eqnarray}\label{3w}
{\bf k}_1={\bf k}_2+{\bf k}_3,\quad{\omega_1}={\omega_2}+{\omega_3},
\end{eqnarray}
where ${\bf k}$ is a wave vector and ${\omega}=\omega({\bf k})$ is the frequency, determined by a dispersion relation [$\omega_j=\omega({\bf k}_j), j=1,2,3].$
The relations (\ref{3w}) correspond to the conservation of momentum and energy:
\begin{eqnarray}\label{ME}
{\bf M}=\int {\bf k}\; N_{\bf k}\; d{\bf k},\quad E=\int \omega({\bf k})\; N_{\bf k}\; d{\bf k},\quad
\end{eqnarray}
where $N_{\bf k}(t)=\varepsilon_{\bf k}/\omega({\bf k})$ is a wave action spectrum (or the number of quasi-particles), $\varepsilon_{\bf k}(t)$ is the energy spectrum.
A natural question: Does there exist another function $\varphi({\bf k})$ --- besides ${\bf k}, \omega({\bf k})$ --- which is also conserved in the interactions (\ref{3w})
\begin{eqnarray}\label{extra}
\varphi_1=\varphi_2+\varphi_3,\quad[\varphi_j=\varphi({\bf k}_j)] \; ?
\end{eqnarray}
 \citep{ZSch0,ZSch}. The function $\varphi({\bf k})$ is supposed to be linear independent of ${\bf k},\,\omega$.
More precisely, does there exist a function $\varphi({\bf k})$ so that equation (\ref{extra}) holds for any vectors 
${\bf k}_1,{\bf k}_2,{\bf k}_3$ bound by the resonance relations (\ref{3w}).

This question is similar \citep{Ferapontov} to the one asked by \citet{Boltz1,Boltz2} about particle collisions in a gas: Does there exist some quantity, which is conserved in the main collisions. The latter are pair collisions; they are similar to four-wave interactions (two particles are {\it in} and two the same particles with new momenta are {\it out}).
If there were such quantity, Boltzmann's kinetic equation would have an additional conservation law (besides the usual ones). \citet{Boltz1,Boltz2} showed that no such quantity exists for classical gas. 

Similar, if there existed function $\varphi$ --- satisfying (\ref{extra}) --- then the wave system would preserve integral
\begin{eqnarray}\label{I}
I=\int \varphi({\bf k})\; N_{\bf k}\; d{\bf k}
\end{eqnarray}
in addition to the momentum and energy. The integral (\ref{I}) is exactly conserved in the three-wave resonance interactions, and therefore, by the wave kinetic equation. However, $I$ is not conserved in non-resonance interactions or in higher-order resonance interactions, that involve more than three waves. In the full weakly nonlinear dynamics, the integral ({\ref{I}) is conserved adiabatically, i.e. approximately over long time. Such integral (or just its kernel $\varphi$) is called ``extra invariant''.

The wave systems with extra invariants are rare.
At present, I know only {\it two} 2-dimensional wave systems that have extra invariants: One is provided by the Kadomtsev-Petviashvili equation \citep{ZSch0,ZSch}. The other is the system of Rossby waves or drift waves in plasma \citep{BNZ,B1991}. 

Let us stress that it is possible to find conservation laws in a nonlinear system by considering only linear dynamics.

The aim of the present paper is to establish extra invariant in MHD of a layer of conducting fluid on a celestial body (planet or star). There is equilibrium, when resting fluid is penetrated by a strong toroidal (zonal, parallel  to the equator) magnetic field. We will see that waves on the background of this equilibrium possess extra invariant (\ref{I}) with unusual kernel
\begin{eqnarray}\label{ArcTan}
\varphi({\bf k})=
\arctan\frac{q+\sqrt{3\alpha}p}{{\sf m}(q^2+\alpha p^2)}
-
\arctan\frac{q-\sqrt{3\alpha}p}{{\sf m}(q^2+\alpha p^2)},\quad
\end{eqnarray}
where ${\bf k}=(p,q)$ is a wave vector ($k^2=p^2+q^2$), and
\begin{eqnarray}\label{rhoAlpha}
{\sf m}=\frac{f A}{\beta c}, \qquad \alpha=\frac{A^2+c^2}{c^2},
\end{eqnarray}
$f$ is the Coriolis parameter, $\beta$ is its latitudinal derivative, $A$ is the Alfv\'{e}n speed (based on the background toroidal magnetic field), and $c$ is the gravity wave speed. The present paper considers dynamics on the $\beta$-plane.
 
This extra invariant is connected to the dynamo phenomenon.
In the second part (Section 3), we will see that the presence of the invariant (\ref{ArcTan}) requres large-scale energy to accumulate in the toroidal magnetic field. This energy supply can maintain -- and even increase --- the background magnetic field $A$. (This would happen on a longer time scale than the typical times of linearized dynamics.) Such feedback on the background magnetic field means dynamo action. However, the present paper does not study this feedback.

In the past \citep{Balk2014,B2022}, I considered a couple of specific situations where the Rossby wave extra invariant emerged. It turns out that a simple modification of the Rossby wave extra invariant can include general dynamics. Besides, the general consideration simplifies the exposition of the extra invariant and makes it more transparent.
In particular, we do not need the quasi-geostrophy (with or without magnetic forces). Neither we need to neglect the Alfv\'en speed $A$ in comparison to the gravity speed $c$.

\section{Extra Invariant}
\label{Sect: Extra}

Let us consider waves on the background of resting fluid with strong toroidal magnetic field. The consideration is possible in two ways,  producing the same result.

First, one could note that the most important is a certain shallow layer,
like Sun's tachocline  \cite[][]{SpieZahn,DikpatiGilman26} or Earth's {\it ocean of the core} \citep{Brag98,Brag07,HardyWong19}. 
The magnethydrodynamics (MHD) of such layers can be studied using the system of ``shallow water'' MHD (SMHD) introduced by \citet{gilman}. 
Linearization of SMHD produces a linear system that describes several wave modes; if the parameters are constant, the waves are determined by the dispersion relation \citep{Schecter01}
\begin{eqnarray}\label{SchecterL}
\omega^4&-&\omega^2(f^2+c^2 k^2+2A^2p^2)\nonumber\\
&+&A^2p^2(c^2k^2+A^2p^2)=0.
\end{eqnarray} 
Here $f$ is the Coriolis parameter; $c=\sqrt{g H_0}$ is the gravity wave speed, $H_0$ is the depth of the shallow layer at rest, $g$ is the reduced gravity \citep[e.g.][]{Va}; $A$ is the Alfv\'en speed, based on the background toroidal magnetic field. 

The same linear system and dispersion relation (\ref{SchecterL}) can be obtained from the linearization of the full 3D MHD. We need to assume the hydro-static approximation [as \citet{Brag98} noted, the gravity is the strongest force, by far, but it acts only in the vertical]. Then the system allows separation of variables, when each of the unknowns is a function of the vertical coordinate, times a function of horizontal  coordinates and time. The geostrophic approximation is not necessary for separation of variables. Now $c^2$ comes as separation constant; it is the eigen-value of some Sturm-Liouville problem. The hydro-static balance seems appropriate, since we aim to consider slow long waves; this balance is also used to derive SMHD.

The equation (\ref{SchecterL}) shows that even small perturbation can drastically alter the frequency of slow long waves, when the last term on the right is extremely small. If we neglect this term, the equation (\ref{SchecterL}) has double root $\omega=0$. 

The small perturbation can take the form of slow latitudinal variation of parameters.
When latitudinal variation of $f$ is taken into account (while $A$ and $c$ are held fixed), the dispersion relation acquires a certain term linear in $\omega$ \citep{ZaqOliBalShe}
\begin{eqnarray}\label{Zaqa}
\omega^4&-&\omega^2(f^2+c^2 k^2+2A^2p^2)-\omega c^2\beta p\nonumber\\
&+&A^2p^2(c^2k^2+A^2p^2)=0.
\end{eqnarray} 

This dispersion relation agrees with the one derived by \citet{Hide66} 
\begin{eqnarray}\label{Hide}
k^2\omega^2-\beta p\omega+A^2p^2 k^2=0
\end{eqnarray} 
for the 2D dynamics of the Taylor-Proudman columns extending through the entire liquid core and constrained by the core-mantle boundary.\footnote{Hide considered background magnetic field with two horizontal components (toroidal and poloidal), but we ignore the poloidal component.}
The meaning of parameter $\beta$ in (\ref{Hide}) is different: This quantity is due to latitudinal variation of the {\it depth} (not variation of the {\it Coriolis force}), so, it has the opposite sign. Hide considered fluid velocity with two components, related by the 2D incomprehensibility condition. This filters out the fast waves, dropping the degree of the dispersion relation from 4 to 2. Besides, this assumes the infinite Rossby radius of deformation, or in other words, $f=0$. To obtain (\ref{Hide}) from (\ref{Zaqa}), we also need to neglect $A^2p^2$ compared to $c^2 k^2$.
Finally, Hide considered nonzero background fluid velocity (to account for latitudinal differential rotation), so that his $\omega$ is the Doppler shifted frequency. The Doppler shift can be also supposed in (\ref{Zaqa}). 

Solving (\ref{Zaqa}), we find the waves with smallest frequency
\begin{eqnarray}\label{0}
\omega=\frac{A^2}{\beta}p (q^2+\alpha p^2)+O(k^5),\quad (k\rightarrow 0).
\end{eqnarray}
It equally follows from either (\ref{Zaqa}) or (\ref{Hide}). The long waves with so small frequency (\ref{0}) are easily excited, even by the weakest forcing. I believe, these waves determine the layer dynamics.

Re-scaling $p\rightarrow p/\sqrt{\alpha}$ reduces (\ref{0}) to the Rossby dispersion relation in the long-wave limit, $\omega=p k^2$, and shows the existence of the extra invariant \citep{BNZ}; in the original variables it is
\begin{eqnarray}\label{0extra}
\varphi=\frac{p^3}{q^2-3\alpha p^2}.
\end{eqnarray}
However, there are two issues with the function (\ref{0extra}):
\begin{enumerate}
\item It has a nonintegrable  singularity. So, the corresponding integral (\ref{I}) is formal and has no physical meaning (it cannot be measured in experiments) \citep[][]{BaYo}.  
\item The function (\ref{0extra}) is not sign-definite. So, it does not restrict fluxes of other conserved quantities through the Fourier space.
\end{enumerate}
Both these issues can be resolved by considering the next approximation when solving the equation (\ref{Zaqa}) 
\begin{eqnarray}\label{Taylor}
\omega=\frac{A^2}{\beta}p (q^2+\alpha p^2)\left[1-{\sf m}^2(q^2+\alpha p^2)\right]+O(k^7),\quad
\end{eqnarray} 
parameters $\alpha$ and ${\sf m}$ are defined in (\ref{rhoAlpha}).
A better approximation with the same asymptotic accuracy $O(k^7)$ is provided by the Pad\'e approximant
\begin{eqnarray}\label{Pade}
\omega=\frac{A^2}{\beta}\frac{p (q^2+\alpha p^2)}{1+{\sf m}^2(q^2+\alpha p^2)}.
\end{eqnarray} 
The Taylor expansion (\ref{Taylor}) quickly separates from the exact lowest frequency (and even changes sign): For large wave numbers, --- if $p\sim q\sim k\rightarrow\infty$ --- the frequency (\ref{Taylor}) grows like 
$\omega\sim k^5$ while according to the dispersion relation (\ref{Zaqa}), $\omega\sim k$. 

The second approximation that follows from solving the dispersion relation (\ref{Hide}) produces a different result, since the equation (\ref{Hide}) has no $f$. Then waves seem to have no extra invariant. We see that stratification is needed for the existence of the extra invariant. The simplest stratification is shallow layer over abyss of heavier fluid \citep{Va}.

In (\ref{Pade}), we again, re-scale $p$ so that $\alpha p^2$ becomes $p^2$. Measuring  $k$ in units ${\sf m}^{-1}$ we transform the frequency (\ref{Pade}) to the Rossby wave dispersion relation 
\begin{eqnarray}\label{Rossby}
\omega=\frac{p k^2}{1+k^2}.
\end{eqnarray}
The waves with this dispersion are known to have extra invariant
\begin{eqnarray}\label{RossbyExtra}
\varphi=\arctan\left(\frac{q+p\sqrt{3}}{k^2}\right)-\arctan\left(\frac{q-p\sqrt{3}}{k^2}\right)
\end{eqnarray} 
\citep{B1991,balk2024extra}. Notice that the dispersion relation (\ref{Rossby}) has $k^2$ in the numerator, unlike the standard Rossby wave dispersion relation. However, this has no effect on the extra conservation, since the standard Rossby wave dispersion relation is $p/(1+k^2)=p-\omega$, and both quantities $p$ and $\omega$ satisfy the resonance conditions (\ref{3w}). Transforming back to the original $p,q,\omega$, we find the function (\ref{ArcTan}). 
 
The parameter ${\sf m}$ can be considered as magnetic analog of the Rossby radius of deformation ${\sf r}=c/f$.

\section{Energy accumulation in the large-scale toroidal magnetic field}
\label{Sect: Energy}

We will see now that the extra invariant (found in the previous section) constrains the energy flux through the scales: It requires the large scale energy to accumulate in toroidal magnetic field.

First,  let us note an obvious fact: Any linear combination of invariants is an invariant.

In particular, the linear combination of the energy and the $x$-component $P={\hat x}\cdot {\bf M}$ of the momentum (\ref{ME})
\begin{eqnarray}\label{enstrophy}
 F=P-\frac{\beta {\sf m}^2}{A^2}\,E=\int \frac{p}{1+{\sf m}^2(q^2+\alpha p^2)}\;N_{\bf k} dp\,dq \quad
\end{eqnarray} 
is conserved. Let us call it {\it enstrophy}. The standard  argument shows that the energy follows the direct cascade towards small scales, while the enstrophy follows the inverse cascade towards large scales.\footnote{Our energy and enstrophy are interchanged compared to the standard situation, because the dispersion relation (\ref{Pade})  additionally contains factor $q^2+\alpha p^2$ in the numerator.}

Nevertheless, the energy accumulates in large scales. Indeed, the enstrophy
carries with it the energy.
So, most of the energy --- supplied by the source --- flows to small scales and dissipates there. But a fraction of supplied energy has to go to large scales. If the source acts for a long time, the energy accumulates in the large scales.  

Using the dimensional considerations, one can derive the Kolmogorov-type energy spectra for the turbulence of waves with dispersion relation (\ref{0}). This spectrum has infrared divergence \citep{Balk2014}, indicating the energy accumulation in large scales. It is actually typical across different physical systems that the large scales carry most of the energy.

Now, using the extra invariant, we will see that the energy accumulates not in any large scales, but specifically in toroidal magnetic field, where $|p|\ll|q|$.

When $p\rightarrow 0$, the function (\ref{ArcTan}) behaves like
\begin{eqnarray}\label{phiAsy}
\varphi=\frac{2{\sf m}\sqrt{3\alpha}\;p}{1+{\sf m}^2 q^2}\;+\;O(p^3).
\end{eqnarray}
So, it is possible to find the linear combination of functions $\varphi(p,q)$, $p$, and $\omega(p,q)$ which is $O(p^3)$ {\it for all values of $q$}. This linear combination is
\begin{eqnarray}\label{tildePhi}
\tilde\varphi=\varphi\;-\;2{\sf m}\sqrt{3\alpha}\;p\;+\;2{\sf m}^3\sqrt{3\alpha}\frac{\beta}{A^2}\;\omega.
\end{eqnarray}
It defines extra invariant
\begin{eqnarray}\label{tildeI}
\tilde I=\int\tilde\varphi({\bf k})\; N_{\bf k}\; dp\,dq=\int\frac{\tilde\varphi}{\omega}\;\,\varepsilon_{\bf k}\;dp\,dq.
\end{eqnarray}
The ratio $\tilde\varphi/\omega$ shows how much extra invariant (\ref{tildeI}) is carried by the energy ``parcel'' 
$\varepsilon_{\bf k} dp dq$.
Figure \ref{fig:ContourPlot} shows the contour lines of the ratio $\tilde\varphi/\omega$ as a function of $p,q$.
\begin{figure}
\includegraphics[width=\columnwidth]{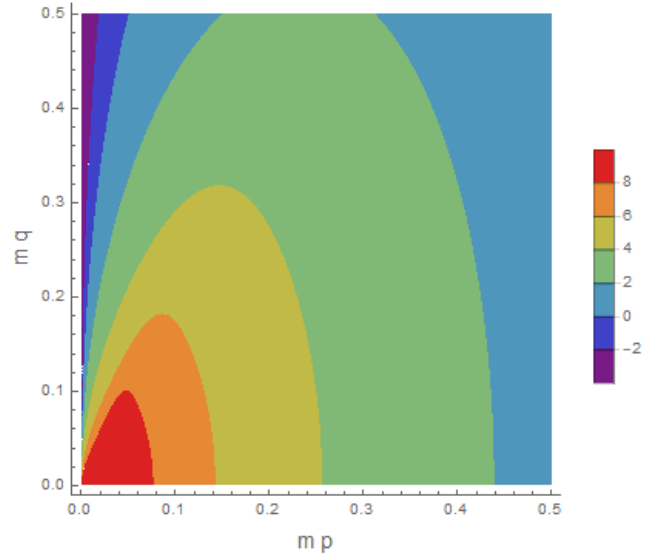}
\caption{Contour plot of the ratio $\tilde\varphi/\omega$ in logarithmic scale: Color-bar shows the values of 
$\ln\left({\tilde\varphi}/{\omega}\right)$. Each of the two extreme areas (where this ratio is bigger than $e^8$ or smaller than $e^{-2}$) is shown by a single color. Since the ratio $\tilde\varphi/\omega$ is even in both $p$ and $q$, the plot is only for positive $p,q$. For this particular figure, $\alpha=2$, i.e. $A=c$. }
\label{fig:ContourPlot}
\end{figure}
Suppose, the source generates energy at some intermediate scale (e.g. $p\sim q\sim k\sim 0.5 {\sf m}^{-1}$).
As the energy is transferred to smaller $k$, it has to accumulate along the $q$ axis, i.e. in the region $|p|\ll |q|$. Indeed, if the energy were to accumulate at large scales away from the region $|p|\ll|q|$, then too much extra invariant would be required. This is because the Fig.\ \ref{fig:ContourPlot} shows large values  of the ratio $\tilde\varphi/\omega$ away from the $q$-axis, much larger than in the source region. Only near the $q$-axis, the values of this ratio are not exceedingly large.

Now, the energy of long waves with dispersion (\ref{Pade}) is mostly {\it magnetic} \citep[e.g.][]{B2022}. So, the energy accumulates in the large-scale toroidal magnetic field.

\begin{subequations}\label{cond}
To come to this conclusion, we used the extra invariant. Its existence follows from the dispersion relation (\ref{Zaqa}) with $f\neq 0$. The extra invariant does not exist in the case of dispersion relation (\ref{Hide}) which has no $f$. Practically, we need that the term $c^2 k^2+2A^2p^2$ in (\ref{Zaqa}) does not dominate $f^2$, at least for long waves participating in the inverse cascade. The longest waves have $k\sim 1/R$, where $R$ is the radius of the fluid layer. Thus, the Rossby radius of deformation ${\sf r}=c/f$ should be much smaller than the radius $R$
\begin{eqnarray}\label{cond1}
{\sf r}\ll R.
\end{eqnarray} 
We have also supposed that the correction of the second approximation (\ref{Taylor}) is smaller than the first approximation, at least for the long waves. This means
\begin{eqnarray}\label{cond2}
{\sf m}\ll R.
\end{eqnarray} 
We have also said that the frequency  (\ref{0}) is the smallest out of all frequencies defined by the dispersion relation (\ref{Zaqa}). In particular, it should be smaller than  the Alfv\'en frequency $\omega=A \, p$, at least for the long waves in the inverse cascade. This means
\begin{eqnarray}\label{cond3}
\ell\ll R,
\end{eqnarray}
where $\ell=\sqrt{A/\beta}$. The third condition (\ref{cond3}) follows from the first two (\ref{cond}ab),
since $\ell=\sqrt{{\sf m}{\sf r}}$.
\end{subequations}
It is unclear if the conditions (\ref{cond}abc) are indeed essential or just technical, used in our derivation. This is because the Rossby wave dispersion relation emerges when $\ell\sim{\sf r}$, without these conditions \citep{B2022}.

For Earth, estimates \citep{B2022} show 
\begin{eqnarray}\label{Earth}
{\sf r}\sim{\sf m}\sim\ell\sim 90 \mbox{km}, 
\end{eqnarray}
while $R\sim 3475$ km, and the conditions (\ref{cond}) are satisfied.

\section{Conclusion}
\label{Sect: Conclusion}

We considered the MHD of conducting fluid layer on a celestial body (planet or star), when along with magnetic forces, the Coriolis and buoyancy forces are also important. The dynamics has simple equilibrium, when fluid is at rest, and magnetic field is toroidal and constant.
We studied the slowest waves on the background of this equilibrium. 

We saw that these waves have extra invariant, conserved adiabatically. The invariant is given by the equations (\ref{I})-(\ref{ArcTan}). Equivalently, the invariant is given by equations (\ref{tildeI}),  (\ref{tildePhi}), and (\ref{ArcTan}). 

We also saw that the presence of the extra invariant leads to the energy accumulation in the large-scale toroidal magnetic field. This energy influx can maintain --- and even increase --- the background toroidal magnetic field. Such feedback would constitute dynamo action. However, the detailed mechanism of this feedback is beyond the present paper. Also, the following important question is left out. The toroidal magnetic field has no signature outside the celestial body. It is unclear whether this toroidal field leads to the magnetic field noticable outside the planet or star.
 
The idea that the dynamo could be a weakly nonlinear phenomenon was suggested long time ago 
\citep[e.g.][]{Hide66,Brag67}. The existence of the extra invariant supports this idea.

\bibliographystyle{apj}
\bibliography{My}

\newpage

\end{document}